\documentclass[a4paper]{article}

\usepackage{INTERSPEECH2022}
\usepackage{xcolor}
\usepackage{hyperref}
\usepackage{url}

\usepackage{etoolbox}
\apptocmd{\sloppy}{\hbadness 10000\relax}{}{}

\title{Accented Speech Recognition: Benchmarking, Pre-training, and Diverse Data}

\name{Al\"ena Aks\"enova\textsuperscript{1},
Zhehuai Chen\textsuperscript{1},
Chung-Cheng Chiu\textsuperscript{2},
Daan van Esch\textsuperscript{1},
Pavel Golik\textsuperscript{1},
Wei Han\textsuperscript{2}, \\
Levi King\textsuperscript{1},
Bhuvana Ramabhadran\textsuperscript{1}, 
Andrew Rosenberg\textsuperscript{1},
Suzan Schwartz\textsuperscript{1}, 
Gary Wang\textsuperscript{1}}
\address{
  Google Speech\textsuperscript{1}, Google Brain\textsuperscript{2}}

\email{\{alenaks,zhehuai,chungchengc,dvanesch,golik,weihan,\\leviking,bhuv,rosenberg,suzan,wgary\}@google.com}

\begin{document}

\maketitle
\begin{abstract}
  Building inclusive speech recognition systems is a crucial step towards developing technologies that speakers of all language varieties can use.
  Therefore, ASR systems must work for everybody independently of the way they speak. 
  To accomplish this goal, there should be available data sets representing language varieties, and also an understanding of model configuration that is the most helpful in achieving robust understanding of all types of speech.
  However, there are not enough data sets for accented speech, and for the ones that are already available, more training approaches need to be explored to improve the quality of accented speech recognition.
  In this paper, we discuss recent progress towards developing more inclusive ASR systems, namely, the importance of building new data sets representing linguistic diversity, and exploring novel training approaches to improve performance for all users.
  We address recent directions within benchmarking ASR systems for accented speech, measure the effects of wav2vec 2.0 pre-training on accented speech recognition, and highlight corpora relevant for diverse ASR evaluations.
\end{abstract}


\section{Introduction}

The goal of Automatic Speech Recognition (ASR) systems is to transcribe speech, allowing voices from speakers of differing accents and language varieties to be understood.
Emerging research aims to address this topic by quantifying the quality of ASR models depending on various parameters (e.g. gender, accent), measuring bias in ASR models, identifying data sets to benchmark quality, and improving model robustness for linguistic diversity.
Moreover, diverse speech recognition has increasingly become the focus of researchers and policy makers. For example, France is considering 
banning discrimination against accents\footnote{\scriptsize \url{https://www.bbc.com/news/world-europe-55069048}}; and some researchers aim to investigate how the bias affecting certain groups of speakers arises in speech recognition models, and how to mitigate this bias \cite{LangBiasEd2022}.

Literature reporting on results of various model evaluations shows that ASR models seem to frequently struggle with accented speech, be it achieving significantly higher word error rate (WER) for non-standard language varieties such as African American Vernacular English \cite{Koenecke7684}, code-switching \cite{vielzeuf2021e2e}, or observing increased WER in certain geographic areas that could be correlated with regional dialects \cite{AksenovaEtAl2020}.
In turn, this can make speakers of certain linguistic varieties feel excluded, and might prompt them to try ``standardizing'' and slowing down their speech \cite{50917}.
This is why measuring bias % \cite{feng2021quantifying} 
and finding ways to improve ASR models given such factors as linguistic diversity, data unavailability, model architecture, and others is so crucial.
An increasing amount of research focuses on finding model configurations or approaches that would help to achieve comparable recognition rates for speakers of different language varieties.
Accented speech is indeed a challenge for ASR, and some research questions the readiness of end-to-end models for industrial use \cite{vielzeuf2021e2e}.
%Additionally, 
ASR has a significant social impact, since its use cases range across multiple applications, including high stakes areas such as medical transcription \cite{mani-etal-2020-towards}.

In our paper, Section 2 presents an overview of the latest methods and challenges within the scope of ASR benchmarking, focusing on recognizing linguistically diverse speech.
Section 3 provides a brief overview of approaches to improving ASR systems for accented speech, emphasizing experiments exploring various types of wav2vec 2.0 pre-training with different types of data, e.g.\ spontaneous unsupervised, synthetic, and accented.
Finally, in Section 4, we outline the current state of accented data sets; discussing those publicly available, and highlighting linguistic aspects insufficiently represented.

\section{Benchmarking ASR systems for accents}

Understanding demographic information is crucial for benchmarking ASR quality for diverse groups of speakers, such as gender, age, speakers of sociolects or regional language varieties, non-native (L2) speakers, people experiencing speech impairments, and more \cite{aksenova-etal-2021-might}.
Indeed, WER on a corpus of African American Vernacular English (AAVE) is sometimes observed to be as much as $85$\% higher than WER for a corpus of Standard American English \cite{Koenecke7684}.
Researchers also show that the quality of speech recognition might not be the same for various genders \cite{LiaoEtAl2015,tatman-2017-gender}. % Tatman2017
A study exploring the link between increased ASR errors and dialectal features demonstrated that geographic areas with increased WER might significantly overlap with regions that have certain dialectal features \cite{AksenovaEtAl2020}.
For example, higher WER scores are observed in the Southern United States, likely due to certain linguistic features of AAVE and Southern American English spoken in those regions.
This  shows that metadata such as sociolect, gender, age, location, native language, and more are determining for a multifaceted analysis of diverse ASR quality. 

%\AR{It might be useful to enumerate how accent can impact speech, and therefore generalization across accents.  E.g. pronunciation differences can limit pronunciation modeling accuracy (if used) or the implicit relationship between acoustic inputs and output graphemic-derived units. Differences to syntax and lexical choice lead to differences the target sequence p(y) whether modeled by an LM or implicitly by an e2e model.  Recording conditions and topic can also differ by corpus making comparisons between, say, Standard American English speech in Switchboard and AAVE speech in CORAAL have different properties due to data collection decisions like audio encoding and style of conversion (spontaneous, prompted phone call, vs. sociological interview).}

\begin{table*}[t]
\begin{center}
\caption{Results with wav2vec 2.0 pre-training and noisy student self-training (NST) on accented data (WER in \%).}
\label{tab:pretrain}
\begin{tabular}{cc|cccc} \toprule
    {Pre-training} & {NST} & {CORAAL:ATL} & {CORAAL:DCB} & {CORAAL:PRV} & {GMU Accent} \\ \midrule\midrule
     - & - & 17.5 & 16.6 & 24.7 & 8.2
  \\ \midrule\midrule
     Libri-Light & - & 16.8 & 17.0 & 22.0 & 5.4 \\\midrule
     YT-U & - & 14.2 & 15.3 & \textbf{20.1} & 4.7   \\\midrule
     YT-U & YT-T & \textbf{12.5} & \textbf{14.0}  & 20.3 & \textbf{4.0}  \\ \bottomrule
\end{tabular}
\end{center}
\vspace{-4mm}
\end{table*}

It is important to point out that accents, dialects and other language varieties are complex linguistic phenomena that span multiple dimensions. Obviously, differences in pronunciations affect the relationship between acoustic input and written output symbols. But dialects can also vary in grammar and vocabulary (a common feature e.g.\ in AAVE or German as spoken outside of Germany), affecting the distribution over the target sequences $p(y)$. This poses a challenge to the language modeling capability, whether in an external LM or one implicitly learned by a transducer model. It is difficult to accurately isolate linguistic features from other aspects affecting ASR quality and control for all variables. Even acoustic conditions such as recording quality and background noise can correlate with geographic regions, which are in turn correlated to demographic differences between populations. Not only will the conversation style (spontaneous, prompted phone calls, scripted media, interviews, etc.) impact recognition accuracy, but also its content. A benchmark that does not address these issues explicitly is likely to confound multiple aspects and mask the real ASR accuracy.

Benchmarking ASR quality when it comes to accented speech is challenging, since multiple parameters should be accounted for: WER for different speech varieties, number of affected speakers, types and severity of misrecognitions, speed, and others.
Importantly, to understand how an ASR model performs for various groups of speakers, there should be a representative data set for every group, and any overall scores should take all individual scores for speaker groups into account. 

One approach to having a deeper insight into ASR quality is to consider a distribution of per-utterance WER scores for the given corpus, and determine the median and the last decile P\textsubscript{90} of the distribution instead of a single WER score \cite{riviere2021asr4real}.
Alternatively, it is possible to use a population-weighted evaluation \cite{aksenova-etal-2021-might} where the overall WER is defined by all WER scores per speaker group normalized by the number of speakers.
In the latter scenario, the number of speakers within that group defines the weight of the WER scores for the corresponding accents.

However, if no transcribed or supervised data for language varieties is available, an alternative approach to testing ASR system fairness would be to use available untranscribed samples.
In this approach, samples from diverse speaker groups undergo various acoustic transformations (e.g.\ noise, frequency scaling, etc.), and the scorer checks how the resulting recognition is affected by those transformations \cite{rajan2022aequevox}.
The overall quality is likely to be lower for a certain accent if these synthetic perturbations significantly worsen the recognition rates of the corresponding accented set.
In this case, the data does not necessarily need to be supervised, since it is the stability of the machine transcript that gives insight into the confusion of the model due to the introduced transformations.
%AR: not a comment to be addressed - but i like this technique.

\section{Improving quality for accented speech}

To find ways of improving ASR quality for all speakers, it is important to study the impact of the training data, training strategies, and model configurations on the evaluations that use accented test sets.
In this section, we present several experiments exploring effects of wav2vec 2.0 pre-training using various types of unsupervised data, and discuss their impact on the recognition of AAVE and non-native (L2) accented speech.

\subsection{Previous work}

While it is possible to achieve higher accuracy with specialized models trained on various accents separately, it is usually more convenient to train a single model on multiple accents simultaneously. Such multi-dialect models tend to be easier to maintain, more robust and require no explicit dialect selection, which may be inaccurate. 
%If the model is trained with the data representing various language varieties, the interesting part is to understand how the information about accents is stored in the neural network's layers. 
%\AR{why is this the interesting part?  the only part of the network that is used is the output unless this information is used to modify the architecture.  And -- unless the network is frozen or otherwise constrained -- architecture changes can lead to different information being stored in different network structures.}
One prominent line of research focuses on understanding which parts of the neural networks process and encode accent information.
For example, research exploring the weights of the hidden layers of an end-to-end system DeepSpeech2 for accented speech shows that the first RNN layer contains the most information about accents \cite{prasad-jyothi-2020-accents}. 
It suggests that this part of an end-to-end model can be adapted to learn abstract representations that are less accent specific.

Other researchers employ an accent classifier to explicitly annotate utterances with certain accent-indicating features that are in turn added to the input of ASR models \cite{JainEtAl2018}.
Modeling approaches include accent conversion \cite{8462258}, where a transformation is applied to a non-native utterance to make it sound as if the speaker had a native accent.
For a thorough overview of research on improving speech recognition for accents, see \cite{RevASRSurvey}.

\begin{table*}[t]
\begin{center}
\caption{Results for adding synthetic data to wav2vec 2.0 pre-training (WER in \%).}\label{tab:tts}
\begin{tabular}{ll|ccc} \toprule
    {Pre-training data} & {Fine-tuning data} & {CORAAL:ATL} & {CORAAL:DCB} & {CORAAL:PRV} \\% & {GMU Accent} \\
    {(unsupervised)} & {(supervised)} & & & \\ \midrule\midrule
    1 M hrs YT &  1000h & 12.8 & 16.5 & \textbf{23.0} % & \textbf{6.0}
  \\ \midrule
    \begin{tabular}[l]{@{}l@{}}$+$\textbf{100k hrs} TTS utterances \end{tabular}  &  1000h & \textbf{12.1} & \textbf{15.8} & 23.4 % & 6.3   
    \\\midrule
\end{tabular}
\end{center}
\vspace{-4mm}
\end{table*}

\subsection{Improving ASR quality via pre-training}

Our experiments show that pre-training of various kinds is able to improve recognition on accented data.
Importantly, it does not harm performance of the same models on the standard test sets. 
Previous work suggests that pre-training is also helpful in settings when little data is available, e.g.\ for building ASR systems for Sub-Saharan African languages \cite{ritchie2022ssa}.
In this section, we consider pre-training of wav2vec~2.0 models \cite{NEURIPS2020_92d1e1eb} using natural and synthetic unsupervised speech.
Our ASR model is a large hybrid autoregressive transducer (HAT)~\cite{variani2020hybrid} with Conformer layers~\cite{gulati2020conformer} in both encoder and prediction network.

\subsubsection{Pre-training with unsupervised spontaneous speech}

In the first set of experiments we explored the effect of wav2vec 2.0 pre-training with spontaneous untranscribed speech \cite{NEURIPS2020_92d1e1eb}.
The settings we considered are pre-training using unsupervised Libri-Light \cite{KahnEtAl2020} and unsupervised YouTube (YT-U) data \cite{zhang2021bigssl}. We also compared the results with a wav2vec 2.0 model without pre-training.
Additionally, we explored the effects of noisy student self-training (NST) \cite{park2020improved} using some of the transcribed public domain YouTube data (YT-T).
%%LK: Next line, maybe: 
In all cases, we trained on the entire SpeechStew data set, totalling roughly 5k hours \cite{chan2021speechstew}.
%In all cases, we trained on 5000 utterances sampled from the SpeechStew data set \cite{chan2021speechstew}.

%%LK: CORAAL utts average 20 seconds; GMU utts average 29 seconds.
We report the results on AAVE data represented by the CORAAL corpus \cite{coraal}, and non-native (L2) accented English data represented by the GMU Accent data set \cite{Weinberger2015}.
AAVE data was collected in Atlanta, GA (ATL, 11 hours), Washington, DC (DCB, 34.5 hours) and Princeville, NC (PRV, 31 hours).
%AAVE data was collected in three different locations: Atlanta, GA (ATL, 2K utterances), Washington, DC (DCB, 6.2K utterances) and Princeville, NC (PRV, 5.6K utterances).
It is important to point out that the PRV data exhibits significantly more diverse dialectal features than sets coming from the other two locations \cite{Koenecke7684}.
%GMU Accent includes data produced by native speakers of ~200 various languages; 2.2K utterances total.
GMU Accent consists of 18 hours of speech produced by native speakers of ~200 various languages.

The results in~\autoref{tab:pretrain} show that the model pre-trained on unsupervised YouTube data together with the NST component gives the overall best results.
The WER on AAVE test sets decreases by 18 to 28\% relative. The pre-training is even more effective on L2 data from the GMU Accent corpus, where the WER drops from 8.2 to 4.0\%.

\subsubsection{Pre-training using synthetic speech}

Another way of sourcing more relevant data is to use synthetic speech, e.g.\ with encoder pre-training following the \textit{tts4pretrain} algorithm for various accents \cite{chen2021injecting}.
In this scenario, a text-to-speech (TTS) system is employed to augment the training data given text.
This approach helps to expand the scope of the available data, enriching it with potentially novel lexical items.

For our experiment, we follow the setup laid out in \cite{chen2021injecting}.
We used a wav2vec 2.0 model pre-trained with 1 million hours of untranscribed speech, followed by TTS to further inject synthetic speech into the pre-training with around 100 million utterances of text, corresponding to roughly 100k hours of synthesized speech. The TTS system is trained on LibriTTS dataset. After pre-training, we fine-tune the system on a small subset of supervised data (around 1000 hours).

The results in~\autoref{tab:tts} show that synthesizing data using TTS and adding it to pre-training decreases the WER on two out of three accented test sets from the CORAAL corpus~\cite{coraal}.
Namely, the model that uses 100M additional synthetic utterances achieves a WER as low as 12.1\% on the AAVE speech.

\subsubsection{Fine-tuning with real accented speech}

If real accented data is available, adding it to the training can be expected to significantly improve the accuracy on accented speech test sets. We find that simply blending in a small amount of in-domain data during supervised fine-tuning can greatly reduce the WER on accented speech while maintaining performance on other test sets. 

For this experiment we sampled accented data from various publicly available sources representing different dialects and language varieties; the accented dataset corpus is roughly 45 hours long.
The non-accented test set that we used for control is composed of short queries sampled from anonymized traffic data.
We evaluated these test sets using pre-trained wav2vec 2.0 models, and observed that the overall quality of recognition improves when accented data is added to the training.
The results in~\autoref{tab:real} show that while the WER on accented data decreases by 30\% (from 28.8\% to 20.4\%), it remains unchanged on the short queries test set.

\begin{table}[t]
\begin{center}
\caption{Fine-tuning with real accented data (WER in \%).}\label{tab:real}
\begin{tabular}{c|cc} \toprule
    {Real accented data} & {Accented} & {Short queries}  \\ \midrule\midrule
    - & 28.8 & 6.7
  \\ \midrule
    +   & \textbf{20.4}   & 6.7  \\ \bottomrule
\end{tabular}
\end{center}
\vspace{-4mm}
\end{table}

\subsection{Further sources of accented speech}

As we showed in the previous subsection, adding various types of unsupervised data to the pre-training improves the recognition of accented speech.
This suggests that further experiments for pre-training are promising, e.g.\ mixing in data from non-target languages in order to achieve better recognition for speakers with non-native accents.

Alternatively, using voice conversion (VC) to produce more training data can be helpful as well.
The idea of a VC model \cite{MOHAMMADI201765} is to perform an on-the-fly conversion from any type of speech to accented speech in the audio domain.
Down the line, ASR models can be jointly trained with this accented data generated by the VC model, potentially augmented with other components, such as consistency loss~\cite{coda}.
Such on-the-fly VC allows us to augment the training data for more accent robust ASR models.
VC is capable of manipulating phonetic and phonological aspects of the accent, and even phonotactic ones depending on the type of VC model.
Overall, VC improves the quality of recognition for accented speech without requiring any additional metadata or human annotations.
However, parameters such as lexical choice and syntactic differences cannot be addressed by VC directly without adding a TTS generator.
Furthermore, recent studies suggest that VC is particularly helpful for improving ASR in very low-resource settings \cite{baas2021voice}.

%\AR{maybe describe what aspects of accent can be manipulated by voice conversion? specifically - lexical choice and syntactic differences cannot be synthesized by VC (but could be by TTS) while phonetic and phonological   differences could be (and maybe phonotactics depending on the VC model).}
% Alena -- done, thanks! PTAL at the paragraph above.

One could use the VC model in order to increase speaker diversity of the training data, allowing various accents to be added to already existing data augmentation.
%, thus diversifying the training sample.

\section{Data sets representing linguistic diversity}

Annotating speech utterances with metadata becomes more and more common for many data sets.
For example, gender annotation is provided for LibriSpeech \cite{PanayotovEtAl2015} speakers, and both gender and age buckets are available for Mozilla Common Voice \cite{ArdilaEtAl2019} utterances, one of the biggest open-source speech data sets.
In this section, we describe multiple data sets and approaches that can be used for diverse model evaluations.

\subsection{Regional language data sets}

TIMIT (LDC93S1) %\cite{GarofoloEtAl1993}
represents $7$ dialects of American English, and there are two versions of the CALLFRIEND corpus of phone conversations: one for Standard American English (LDC96S46)
% \cite{CanavanZipperlen1997}
and one for Southern American English (LDC96S47).
% \cite{CanavanZipperlen1997south}
As for dialects spoken in the UK and Ireland, they are represented within an OpenSLR data set (SLR83) \cite{demirsahin-etal-2020-open}. Regional corpora are available for other languages as well, e.g.\ Swiss German \cite{samardzic-etal-2016-archimob}, Galician Spanish \cite{BarcalaEtAl2018} and Arabic dialects \cite{lounnas-etal-2019-building}, along with several multilingual corpora \cite{MediaParl}.
Additionally, public challenges such as \textit{The MGB Challenge} target evaluations of various ASR topics including dialect identification.

Still there are not many data sets that represent speech on a more detailed level of geographical granularity, e.g.\ in various metro areas or regions for a given language.
However, the practice of utterance annotation with at least country-level metadata is becoming more common.
A recent study suggests ways to create such open-source data sets using public domain videos \cite{Coats2019}, showing that the construction of regional corpora can be partially automated.
Other research focuses on machine learning methods to identify dialectal utterances, using the example of local accents spoken in Pakistan \cite{MOHAMMADI201765}.

\subsection{L2 accented sets}

A number of corpora represent accented English speech, including GMU Accent \cite{Weinberger2015}, IDEA \cite{ideaweb}, and ALLSSTAR \cite{allstarweb}.
In GMU Accent corpus, for example, as many as $200$ native languages of the speakers are represented.
Data is available in other languages as well: for example, West Point developed corpora representing both native and non-native speakers for Spanish \cite{LDCspanish2006}, Russian \cite{LDCrussian2003}, and Arabic \cite{LDCarabic2002}. A European consortium published IFCASL \cite{trouvain-etal-2016-ifcasl}, a corpus of French and German native and non-native speech.

\subsection{Sociolectal data sets}

While it is often possible to find a representation of regional and accented speech in widely spoken languages, less data is available for sociolects (e.g.\ Chicano English or Cockney English).
African American Vernacular English (AAVE), for example, is spoken by approximately 40 million speakers, and is represented within the CORAAL \cite{coraal} and Voices of California %\cite{vocwebsite}
corpora. CORAAL also showcases regional variation within AAVE, comparing data from six metro areas.
However, more work is needed to adequately represent other sociolects.

\subsection{Sets representing speech impairments}

Another category of data sets that we would like to highlight exemplifies various types of speech disorders.
Speech recognition for impaired speakers is a different problem than improving accuracy for various accents. However, this group of speakers might benefit from more available data and training approaches promoting accent robustness as well.
For example, the Euphonia project involved collecting one million utterances representing different types and severities of speech impairments \cite{macdonald21_interspeech}.
More of such data is available in the Whitaker corpus \cite{DellerEtAl1993}, the Dysarthric Speech Database for Universal Access Research \cite{KimHPGHWF08} and EasyCall \cite{turrisi2021easycall}.
A Spanish corpus published in \cite{SazEtAl2008} contains recordings of children with speech impairments in particular.

% \AR{These are orthogonal to accent. If it's important to keep this section maybe consider framing it as a different problem that can be addressed with similar solutions?  benchmarking and more data is needed, but training approaches to promote accent robustness may be applicable to improving performance on dysarthric speech.}
% Alena -- explained, added the second sentence in that paragraph.

\medskip

A growing number of researchers are working towards creating data sets representing different types of factors affecting ASR quality, e.g.\ accented speech, noise, various acoustic environments and other aspects.
For example, ASR4REAL \cite{riviere2021asr4real} is a combination of benchmarks assembled at FAIR that is intended to represent various accents and real-life conditions.

\section{Discussion}

To understand how well ASR models serve people speaking different dialects and accents, we need data representing various language varieties, and a methodology of how to approach the evaluation.
In this paper, we discuss several experiments aiming to improve ASR quality for accented speakers, along with some recent results in the area of ASR benchmarking.

% : using diverse corpora and considering the median or the distribution of WER scores, weighting WER depending on the size of the speaker group that is represented by that data, or see how robust the accented data is when it undergoes various transformations.

Indeed, a crucial research direction is to improve ASR quality for accented speakers.
This goal is impossible to achieve without representative test sets and established benchmarks, since improving recognition for widely used corpora representing mainstream types of speech does not guarantee improving ASR for other groups of accented speakers.
And indeed, without a proper way to measure speech recognition quality for various accents there is no way to know how certain configurations affect different groups of speakers and types of language diversity.
To help researchers with the challenge of obtaining representative data samples for various regions, non-native accents, sociolects, and speech disorders, we provided a quick survey of currently available corpora suitable for evaluating ASR quality on such speech varieties.

%\AR{maybe stress here, and in the introduction, that this is *impossible* without good test sets and established benchmarks.  Many approaches "hope" that a rising tide will lift all ships -- improvement to the most common accent will result in improved recognition of across all accents. But without measurement there is no way to know whether this expectation is realized.}
% Alena -- done, split paragraphs and discussing it now above.

Previous works suggested accent conversion or employing additional accent classifiers to create more in-domain data.
In this paper, we showed how various types of wav2vec 2.0 pre-training and fine-tuning affect the recognition accuracy on accented test sets representing AAVE and non-native speakers. Namely, we explored the effects of pre-training using unsupervised spontaneous and synthetic speech, showing that it indeed helps to achieve higher recognition quality for accents without increasing the WER on the other sets.

To conclude, we would like to highlight the importance of collaboration in this emerging field of inclusive ASR technologies.
Researchers and developers can get insights into challenging areas and language aspects from linguists and dialectologists, who, in turn, could help to analyze the patterns of recognition issues, and address the issues of missing data for some groups of speakers.
Crowdsourcing and community involvement allows for scaling data collection projects, and extensive metadata annotation enables understanding of the overall recognition quality with respect to the factors of language variation.
Finally, it is crucially important for the researchers to share their results and findings, drawing greater attention to this area.

\section{Acknowledgments}

We are grateful for the resources provided by Google, and for our collaborators from Speech and Brain teams, especially Françoise Beaufays, James Flynn, Bo Li, Pedro J.\ Moreno, Daniel Park, Sandy Ritchie, Pierric Sans, Benyah Shaparenko, and Yu Zhang.
For critical work on data sets, our thanks go to Landis Baker, Jonathan Endale, Mandy Jordan, Prescott Nicoll, and Travis Trekell.

\bibliographystyle{IEEEtran}

\bibliography{main}

\end{document}